\documentclass[aip,jcp,reprint,showkeys]{revtex4-1}
\usepackage{graphicx,dcolumn,bm,xcolor,microtype,hyperref,multirow,amsmath,wasysym,amssymb,amsfonts,physics,mathpazo,float,ulem}
\usepackage[version=4]{mhchem}
\usepackage{adjustbox}
\usepackage[latin1]{inputenc}
\usepackage[T1]{fontenc}
\usepackage{lipsum}
\usepackage{caption}

\begin{document}
\title{Evaluating two-electron-repulsion integrals over 
arbitrary orbitals using Zero Variance Monte Carlo: Application to Full Configuration Interaction 
calculations with Slater-type orbitals }

\author{Michel Caffarel}%
\affiliation{Laboratoire de Chimie et Physique Quantiques (CNRS 5626), IRSAMC, Universit\'e P. Sabatier, Toulouse (France)}

\begin{abstract}
A Monte Carlo method for evaluating multi-center two-electron-repulsion integrals 
over any type of orbitals (Slater, Sturmian, finite-range, numerical, etc.)
is presented. The approach is based on a simple and universal (orbital-independent) 
gaussian sampling of the two-electron configuration space and on the use of 
efficient zero-variance Monte Carlo estimators. Quite remarkably, it is shown that 
the high level of accuracy 
required on two-electron integrals to make Hartree-Fock (HF) and configuration interaction (CI)
calculations feasible can be achieved. A first zero-variance estimator is built by introducing 
a gaussian approximation of the orbitals and by evaluating the two-electron integrals 
using a correlated sampling scheme for the difference between exact and approximate orbitals. 
A second one is based on the introduction of a general coordinate transformation.
The price to pay for this simple and general Monte Carlo scheme is the 
high computational cost required. However, we argue that the great simplicity of the algorithm, 
its embarrassingly parallel nature, its ideal adaptation 
to modern computational platforms and, most importantly, the possibility of 
using more compact and physically meaningful basis sets  
make nevertheless the method attractive. 
HF and near full CI (FCI) 
calculations using Slater-type orbitals (STO) are 
reported for Be, CH$_4$ and [H$_2$N(CH)NH$_2$]$^+$ (a simple model of cyanine). 
To the best of our knowledge, our largest FCI calculation involving
18 active electrons distributed among 90 orbitals for the cyanine molecule,
is the most extensive molecular calculation performed so far using pure STO 
orbitals (no gaussian approximation, even for the challenging four-center two-electron integrals).
\end{abstract}
\maketitle

\section{Introduction}
In recent years most of the standard methods of quantum chemistry have been 
revisited within the framework of stochastic processes. In short, the very same equations 
and quantities are considered but, instead of solving the equations 
using standard linear algebra techniques (diagonalization) or explicit calculations 
of very large sums (perturbational quantities), stochastic implementations are employed.
Let us cite the stochastic versions of the second-order M\o ller-Plesset 
(MP2),\cite{mp2} coupled-cluster with single and double and perturbative triple excitations (CCSDT),\cite{thom} complete active space self-consistent field (CASSCF),\cite{manni} multi-reference 
with second-order perturbation (MRPT2), \cite{garniron,sharma}, random phase approximation 
(RPA)\cite{rpa}, GW\cite{gw}, and FCI\cite{alavi} approaches.
In practice, by avoiding the practical limitations in terms of memory (no storage of 
very large vectors and matrices) and 
number of determinants to consider (only a small subspace consisting of 
the determinants contributing the most to the averages is sampled) calculations beyond the limits 
of the standard ''deterministic'' versions 
can be performed. As a representative example, let us mention the recent stochastic CASSCF 
calculation of Smith {\it et al.} involving 44 electrons distributed among 44 active orbitals 
for a model complex of Fe-porphyrin.\cite{mussard} 
From a general perspective, the major driving force behind the active developement 
of stochastic techniques is their very good adaptation to massive parallelism 
and to modern computational platforms
(simplicity of the algorithm, low-memory fingerprint, easy implementation on 
graphics processing units (GPU) and efficient
arithmetic co-processors, cache optimization, etc.).

In the same spirit, we propose here to calculate
the two-electron-repulsion integrals 
of quantum chemistry using a stochastic approach.
As well-known, the choice of the basis functions (orbitals) in electronic structure wave-function 
calculations is one of the critical aspects. Ideally, orbitals should obey 
the electron-nucleus cusp condition removing the divergence of 
the one-electron component of the (local) energy at short electron-nucleus distances; they should 
also display the physically correct exponential-like decay at large distances, and be 
flexible enough to reproduce any type of 
behavior at intermediate distances. Unfortunately, the high computational cost required
to evaluate the very large numbers of integrals involved in calculations limits in practice 
the type of orbitals that can be employed. As well-known, the compromise between
cost and efficiency adopted in virtually all
calculations for molecular systems consists in 
using Gaussian-type basis functions.
Although fast and efficient algorithms have been developed over the years to calcutate 
gaussian integrals, the price to pay 
is the need of using large sets of basis functions, larger than those based on 
more physical representations (for example, Slater-type orbitals 
with the correct cusp and long-range behavior).
Considering the sharp increase of the computational cost of accurate post-HF
methods with the number of basis functions [{\it e.g.}, $N_b^7$ scaling for
the ''gold standard'' CCSD(T), $N_b$ number of basis functions], 
to have the possibility of using more compact basis set is important,
particularly for large systems. 

Here, we present a Monte Carlo approach to calculate
two-electron integrals for arbitrary orbitals (STO, numerical, finite-range, etc.). 
In this approach no analytic integration is performed and only 
the values of the orbitals at each Monte Carlo configuration are to be calculated, 
making the approach particularly simple and general. However, at first sight
using a Monte Carlo approach to calculate accurately low (six)-dimensional integrals 
may appear unrealistic. Indeed, the statistical error is usually large and 
its very slow decay 
with the number $N$ of drawings -$\sim 1/\sqrt{N}$- precludes any brute 
force approach ({\it i.e.} increasing $N$ indefinitely) to improve the accuracy.
Here, this problem is particularly 
acute since a high accuracy on the 
two-electron integrals is known to be needed to get stabilized and unbiaised
HF or post-HF calculations. For example, the use of single precision floating point 
representation is in general not sufficient and an absolute error at least smaller than 
$10^{-8}$ is necessary.\cite{dp}

In this work it is shown that by resorting to a zero-variance strategy
the statistical error on two-electron integrals can be tremendously reduced and 
the targeted accuracy can be attained.
For example, in the case of the biggest system treated here (the cyanine molecule)
an average absolute error of about $\sim 2 \times 10^{-9}$ on the two-electron integrals 
is achieved. From a general perspective, a zero-variance strategy is based on 
the introduction of improved estimators 
having the same average as the standard estimator but a (much) 
smaller variance.\cite{prl_zv} In this way, for a given number of Monte Carlo configurations
(much) more accurate averages (smaller statistical error) can be obtained at essentially the 
same computational cost. When building up such improved estimators it is usually possible 
to define the ideal zero-variance limit where statistical fluctuations entirely vanish.
In practice, approaching this limit is a guarantee of decreasing the statistical error.
In this work on computing two-electron integrals, we define a first zero-variance estimator 
based on a gaussian approximation of the orbitals and on the evaluation 
of the exact integrals using a correlated sampling scheme for the difference between exact
and approximate orbitals. It is most important to emphasize that, 
although a gaussian approximation for the orbitals is 
introduced, the calculated integrals are {\it independent} of this approximation, 
only the magnitude of the statistical error is affected. The zero-variance limit is 
attained in the limit of an exact representation (infinite number of gaussian functions). 
A second zero-variance estimator defined here is obtained 
by introducing a coordinate transformation. In this case it is possible to write down 
a so-called zero-variance equation defining the best transformation. 
In practice, searching for good approximations of this equation is a precious guide to build 
efficient improved estimators. However, once again, we note that the results 
are independent of the quality of the approximation made for 
the transformation. The introduction of zero-variance estimators being instrumental to 
the success of the method, the approach will be referred to as
zero-variance Monte Carlo (ZVMC).

In this work ZVMC is applied to the calculation of two-electron integrals
over Slater-type orbitals. The problem of computing such integrals has a long history 
from the very start of quantum chemistry and
has given rise to numerous works (for references see, {\it e.g.}, [\onlinecite{hoggan,smiles}]). 
Here, it is shown that STO integrals can be 
computed with sufficient accuracy to allow converged HF and FCI-type calculations 
for Be, CH$_4$, and [H$_2$N(CH)NH$_2$]$^+$ (a simple model of cyanine).
To the best of our knowledge, the FCI calculation presented here for the cyanine molecule
involving 18 active electrons (and 6 frozen core electrons) distributed among 90 orbitals is the most extensive molecular calculation performed so far using pure STO orbitals 
(no approximate gaussian expansion for two-electron integrals, even for the challenging four-center
integrals).
However, the price to pay for this simple and general approach is the need 
of using (very) large Monte Carlo statistics. It is clearly the major drawback of the approach. 
However, as illustrated and discussed in this work we believe 
that the unique features of the approach 
nevertheless make the method attractive. 

Finally, let us note that we shall here restrict ourselves to atomic orbitals.
However, there is no fundamental difficulty to consider molecular orbitals; 
this will be presented in a forthcoming work.

The paper is organized as follows. In Sec. \ref{theory} the basic theory is presented.
Some illustrative numerical applications are discussed in section \ref{num}. We first 
present the main aspects of the method with calculations of several 
representative four-center two-electron integrals over Slater-type orbitals. 
Then, HF and near-FCI
calculations are presented for Be, CH$_4$ and [H$_2$N(CH)NH$_2$]$^+$ (a simple model of cyanine).
Finally, a summary and discussion are given in Sec.\ref{conclusion}

\section{General theory}
\label{theory}
In this work we are concerned with the 
calculation of general two-electron-repulsion integrals of the form
\begin{equation}
I= (ab|cd)= 
\int d{\bf r}_1 d{\bf r}_2 
{\phi_a({\bf r}_1) \phi_b({\bf r}_1) \frac{1}{r_{12}} \phi_c({\bf r}_2) \phi_d({\bf r}_2)}
\end{equation}
where the $\phi$'s are real atomic orbitals written under the general unnormalized cartesian form
\begin{equation}
\phi_a({\bf r})= (x-A_x)^{a_x} (y-A_y)^{a_y} (z-A_z)^{a_z} u_a(|{\bf r}-{\bf A}|).
\label{defphi}
\end{equation}
Here, ${\bf A}=(A_x,A_y,A_z)$ is the center, 
${\bf a}=(a_x,a_y,a_z)$ a triplet of non-negative integers (angular momentum vector),
and $u_a(r)$  some general radial part.
Standard examples of radial parts are, {\it e.g.}, $u_a(r) = e^{-\alpha r^2}$ 
for Gaussian-type orbitals (GTO), $u_a(r) = r^{n-l-1} 
e^{-\alpha r}$ for Slater-type orbitals (STO),
or $u_a(r) =F(r) e^{-\alpha r}$ for Sturmian orbitals (where $F$ is the confluent 
hypergeometric function).
Alternatively, the radial part may be defined on a 
one-dimensional grid or using a spline representation. 
Note that the expression for the radial function 
does not need to be the same among orbitals, so mixed basis sets ({\it e.g.} STO-GTO) 
can also be used. In what follows, we will characterize the radial extension of a general orbital
$\phi_a$ by introducing an effective exponent $\alpha > 0$ equal to the inverse of 
the average radial width of the orbital, that is
\begin{equation}
\alpha \sim \frac{1}{\langle r \rangle}_{\phi_a},
\label{definv}
\end{equation}
where ${\langle r \rangle}_{\phi_a}= \frac{\int r^2 dr \;r \;{\phi^2_a}}{\int r^2 dr {\phi^2_a}}$. The exponents associated with $\phi_b$, $\phi_c$, and $\phi_d$ will be denoted as 
$\beta$,$\gamma$, and $\delta$, respectively.\\

The densities $\rho_{ab}({\bf r})$ are defined as
\begin{equation}
\rho_{ab}({\bf r}) =\phi_a({\bf r})\phi_b({\bf r}),
\end{equation}
and the integral writes
\begin{equation}
I=
\int d{\bf r}_1 d{\bf r}_2
\frac{1}{r_{12}} \rho_{ab}({\bf r}_1)\rho_{cd}({\bf r}_2).
\end{equation}
{\it Remark on notation}: 
For simplicity the $abdc$-dependency of the integral has not been indicated here;
in what follows it will be the case for all quantities for which this dependency is obvious, 
except when some confusion is possible.

\subsection{Simple Monte Carlo estimator}
The two-electron-repulsion integral is expressed as
\begin{equation}
I= \int d{\bf r}_1 d{\bf r}_2 
\pi_{ab|cd}({\bf r}_1,{\bf r}_2)\Big[ \frac{1}{r_{12}}
 \frac{ \rho_{ab}({\bf r}_1)\rho_{cd}({\bf r}_2)}{\pi_{ab|cd}({\bf r}_1,{\bf r}_2)}\Big],
\end{equation}
where $\pi_{ab|cd}$ is some arbitrary probability density ($\pi_{ab|cd} \ge 0$ 
and $\int d{\bf r}_1 d{\bf r}_2 \pi_{ab|cd}=1$).
Here, we use a simple factorized gaussian density reproducing
the overall shape of the one-electron distributions, for example
\begin{equation}
\pi_{ab|cd}( {\bf r}_1,{\bf r}_2)= {\Big(
\frac{\sqrt{\zeta \eta}} {2\pi}
\Big)}^3
e^{-\frac{\zeta}{2} ({\bf r}_1-{\bf P} )^2}
e^{-\frac{\eta}{2}  ({\bf r}_2-{\bf Q} )^2},
\end{equation}
with 
\begin{equation}
\zeta= \alpha+\beta\;\;\;\;\;\;\;\;\; \;\;\; \eta= \gamma+\delta
\label{zeta}
\end{equation}
and
\begin{equation}
{\bf P}=\frac{\alpha {\bf A} + \beta {\bf B}}{\alpha+\beta} \;\;\;\;\;\;\;\;\; \;\;\;
{\bf Q}=\frac{\gamma {\bf C} + \delta {\bf D}}{\gamma+\delta}
\end{equation}

After simple changes of variables and relabelling, 
the integral can be written under the form
\begin{equation}
I= \int d{\bf r}_1 d{\bf r}_2
\pi_{0}({\bf r}_1,{\bf r}_2) F({\bf r}_1,{\bf r}_2)
\end{equation}
with
\begin{equation}
F({\bf r}_1,{\bf r}_2) = 
(\zeta \eta)^{-\frac{3}{2}}
\frac{1}{|{\bf u}_1-{\bf u}_2|}
\frac{ \rho_{ab}({\bf u}_1)\rho_{cd}({\bf u}_2)} {\pi_0({\bf r}_1,{\bf r}_2)},
\end{equation}
and
$$
\left\{
\begin{array}{l}
{\bf u}_1 = {\zeta}^{-\frac{1}{2}}{\bf r}_1+ {\bf P}\\
{\bf u}_2 = {\eta}^{-\frac{1}{2}} {\bf r}_2+ {\bf Q}\\
\end{array}
\right.
$$
Here, $\pi_0$ is the product of the normal distribution for each electron coordinate
\begin{equation}
\pi_0({\bf r}_1,{\bf r}_2)= 
\frac{1} {(2\pi)^3}
e^{-\frac{1}{2} ( {\bf r}^2_1 + {\bf r}^2_2 )}.
\end{equation}

To apply Monte Carlo techniques, the integral is rewritten as 
\begin{equation}
I = \langle F \rangle_{\pi_0},
\end{equation}
where $\langle...\rangle_{\pi_0}$ denotes the average over the probability distribution $\pi_0$.
In practice, the integral is evaluated from a finite random sample of $N$
configurations $({\bf r}^i_1,{\bf r}^i_2)$ drawn with $\pi_0$,
\begin{equation}
I \simeq I_N=\frac{1}{N} \sum_{i=1}^N F({\bf r}^i_1,{\bf r}^i_2),
\end{equation}
the exact value being obtained as $N$ goes to infinity.
At finite $N$, the statistical error bar on $I_N$ is calculated using
elementary statistical techniques.

\subsection{Zero Variance Monte Carlo estimators}
The general idea of variance reduction techniques\cite{prl_zv} is to replace 
the initial estimator $F$, by a new ``improved'' one, denoted here as $\tilde{F}$, 
having the same average but a smaller 
variance, $\sigma^2(\tilde{F})$
\begin{equation}
\langle \tilde{F} \rangle = \langle {F} \rangle \;\;\; {\rm with}\;\;\; \sigma^2(\tilde{F}) <
\sigma^2(F).
\end{equation}
In this formula the average is defined over some general probability density and the variance is given by
\begin{equation}
\sigma^2(F)= \langle F^2 \rangle-\langle F \rangle^2.
\end{equation}
As long as the calculation of $\tilde{F}$ is not too expensive, calculating the average 
using $\tilde{F}$ instead of $F$ leads to a decrease of the statistical 
error, the gain in computational cost being essentially proportional to the reduction 
in variance. The ideal zero-variance limit where the
statistical fluctuations vanish is reached 
when $\tilde{F}$ can be made constant for all configurations, more precisely
\begin{equation}
\tilde{F}= \langle \tilde{F} \rangle.
\end{equation}

\subsubsection{Zero Variance using control variates}
The first zero-variance (ZV) estimator introduced here is based on the use 
of the so-called control variate method, {\it e.g.} [\onlinecite{control}]. 
Denoting $F_0$ some approximation of $F$ {\it whose 
average, $\langle F_0 \rangle$, is known} the following improved estimator is considered
\begin{equation}
\tilde{F}= F + \lambda (F_0-\langle F_0 \rangle)
\end{equation}
where $\lambda$ is some real parameter.
By construction, $\langle \tilde{F} \rangle= \langle {F} \rangle$, for all 
$\lambda$.
Minimizing the variance with respect to $\lambda$, the variance of the
optimized estimator is found to be  
\begin{equation}
\sigma^2(\tilde{F})=  \sigma^2({F}) 
-\frac{ {\langle  (F-\langle F \rangle)
                  (F_0-\langle F_0 \rangle) \rangle}^2 
       }
{ \sigma^2(F_0) }.
\end{equation}
As seen, by using the control variate $F_0$ a 
systematic decrease of the variance is obtained, whatever 
the choice of $F_0$. However, a significant variance reduction is possible in practice 
only if the fluctuations of $F_0$ are correlated enough to those 
of $F$, that is, if the correlator 
$\langle (F-\langle F \rangle)(F_0-\langle F_0 \rangle) \rangle$ is large enough.\\

Here, the control variate $F_0$ is chosen
by using some gaussian approximation $\rho^G$ of the exact density $\rho$, more precisely
\begin{equation}
F_0({\bf r}_1,{\bf r}_2) =
({\zeta \eta})^{-\frac{3}{2}}
\frac{1}{|{\bf u}_1-{\bf u}_2|}
\frac{ \rho^G_{ab}({\bf u}_1)\rho^G_{cd}({\bf u}_2)} {\pi_0({\bf r}_1,{\bf r}_2)}.
\end{equation}
The average of $F_0$ given by
\begin{equation}
\langle F_0 \rangle = I^G=
\int d{\bf r}_1 d{\bf r}_2 
\frac{1}{r_{12}} \rho^G_{ab}({\bf r}_1)\rho^G_{cd}({\bf r}_2)
\end{equation}
can be efficiently evaluated using standard algorithms for gaussian integrals.
We now decompose $I$ as
\begin{equation}
I= I^G + \Delta I
\end{equation}
where $\Delta I$ is a residual integral given as
\begin{equation}
\Delta I =\int d{\bf r}_1 d{\bf r}_2 
{\pi_0({\bf r}_1,{\bf r}_2)}
\Delta F ({\bf r}_1,{\bf r}_2)
\label{def_di}
\end{equation}
where
$$
\Delta F({\bf r}_1,{\bf r}_2)
= { ({\zeta \eta })^{-\frac{3}{2}} }
\frac{1}{|{\bf u}_1-{\bf u}_2|}
$$
\begin{equation}
\times
\frac{ 
\rho_{ab}({\bf u}_1)\rho_{cd}({\bf u}_2)- 
\rho^G_{ab}({\bf u}_1)\rho^G_{cd}({\bf u}_2)}
{\pi_0({\bf r}_1,{\bf r}_2)}
.
\end{equation}
The formula for the integral becomes
\begin{equation}
I= I^G + \langle \Delta F \rangle_{\pi_0}
\label{def1}
\end{equation}
where the first 
contribution, $I^G$, is calculated deterministically 
and the residual integral, $\Delta I$, computed with Monte Carlo.
While $\rho^G$ is approaching $\rho$, $\Delta I$ becomes smaller and smaller 
and the same for the statistical error. In the zero-variance limit where $\rho=\rho^G$, the error entirely vanishes. In practice, 
using accurate gaussian approximation leads to (very) important reduction in 
statistical fluctuations.\\

Now, since the integrals are {\it independent} of $\rho^G$ 
-whatever the quality of the approximation- we have a great freedom in choosing the 
way the densities $\rho_{ab}$ are approximated. For example, it can be done 
by using density fitting or related techniques where auxiliary basis sets 
are introduced to approximate products of one-electron functions.
Here, we shall not elaborate on this aspect (this is let for future work) but use instead
the simple procedure consisting in building $\rho^G$ as the product 
of some gaussian approximation $\phi^G_a$ for the orbitals
\begin{equation}
\phi^G_a({\bf r})= (x-A_x)^{a_x} (y-A_y)^{a_y} (z-A_z)^{a_z} u^G_a(|{\bf r}-{\bf A}|).
\label{defphig}
\end{equation}
with
\begin{equation}
u^G_a(r) = \sum_{i=1}^{n_g} c^a_i  r^{2n_i}e^{-\gamma^a_i r^2}.
\label{defua}
\end{equation}
Here, $n_g$ is the number of elementary gaussian functions used, $\{n_i\}$ a fixed set 
of positive integers, and ($c^a_i$, $\gamma^a_i$) the parameters 
resulting from some fitting process, for example by minimizing the $\chi^2$ quantity
\begin{equation}
\chi^2= \int_{0}^{+\infty} r ^2 dr\;\; \frac{1}{r} {[u_a(r)-u^G_a(r)]}^2.
\label{chi2}
\end{equation}

\subsubsection{Zero Variance using a coordinate transformation}
Our second zero-variance estimator is based on the fact 
that a coordinate transformation can 
be used to reduce the statistical error. Let us note 
$[\tilde{\bf r}_1({\bf r}_1,{\bf r}_2),\tilde{\bf r}_2({\bf r}_1,{\bf r}_2)]$
such a one-to-one correspondance. 
The residual integral, $\Delta I$, computed with Monte Carlo , Eq.(\ref{def_di}) writes
$$
\Delta I =\int d{\tilde{\bf r}_1} d{\tilde{\bf r}_2}  
\pi_0(\tilde{\bf r}_1,\tilde{\bf r}_2)
\Delta F({\tilde{\bf r}_1},{\tilde{\bf r}_2})
$$
\begin{eqnarray*}
= 
\int d{\bf r}_1 d{\bf r}_2 \pi_0({\bf r}_1,{\bf r}_2)  \widetilde{\Delta F}({\bf r}_1,{\bf r}_2)
\end{eqnarray*}
with
\begin{equation}
\widetilde{\Delta F}({\bf r}_1,{\bf r}_2)= 
\frac{
\pi_0(\tilde{\bf r}_1,\tilde{\bf r}_2) }
{\pi_0({\bf r}_1,{\bf r}_2)} J({\bf r}_1,{\bf r}_2)
\Delta F(\tilde{\bf r}_1,\tilde{\bf r}_2),
\end{equation}
where $J$ is the Jacobian of the transformation
\begin{equation}
J({\bf r}_1,{\bf r}_2) = \left|
{\rm det} \frac{\partial {\tilde {\bf r}}_\mu}{\partial {\bf r}_\nu} \right| \;\;\;\mu,\nu =1,2
\end{equation}
The expression for the complete two-electron integral thus writes
\begin{equation}
I = I^G + \int d{\bf r}_1 d{\bf r}_2 \pi_0( {\bf r}_1,{\bf r}_2 ) \widetilde{\Delta F}({\bf r}_1,{\bf r}_2).
\end{equation}
or, equivalently
\begin{equation}
I = I^G + \langle \widetilde{\Delta F} \rangle_{\pi_0}.
\end{equation}
Although the Monte Carlo average 
is independent on the transformation, $\langle \widetilde{\Delta F} \rangle_{\pi_0} = 
\langle \widetilde{\Delta F} \rangle_{\pi_0}$, it is not at all true for its variance. Now, 
a precious guide to construct a coordinate-transformation 
leading to a large reduction in variance consists in invoking
the zero-variance equation that 
the ideal transformation (no statistical fluctuations) obeys.
This equation is obtained by equating the quantity to be averaged
to its average,\cite{prl_zv} that is, here
\begin{multline}
\widetilde{\Delta F}({\bf r}_1,{\bf r}_2)=
\frac{
\pi_0(\tilde{\bf r}_1,\tilde{\bf r}_2) }
{\pi_0({\bf r}_1,{\bf r}_2)} J({\bf r}_1,{\bf r}_2)
\Delta F(\tilde{\bf r}_1,\tilde{\bf r}_2)
= \Delta I
\end{multline}
In the next section it will be illustrated 
how this ZV equation can be exploited in the particular case of STO integrals.
Note that, although $\Delta I$ -the unknown quantity to be computed- is present 
in the equation, it is not a problem in practice. Indeed, a simple solution consists
in replacing the exact value $\Delta I$ 
by some approximate one. It is legitimate
as long as the variation of $\widetilde{\Delta F}$
in configuration space -measured for example 
by its variance, is larger than the error made for $\Delta I$,
which is always the case except for very simple cases. 
Once a functional form for the coordinate transformation has been chosen, 
its parameters can be optimized by minimizing the variance of 
$\widetilde{\Delta F}$ evaluated over a fixed set of configurations drawn according to 
$\pi_0$.

\subsection{The case of Slater-type orbitals}
In this section, we make more explicit the general approach just described 
for the important case of STO atomic 
orbitals. The real cartesian unnormalized STO orbitals $\phi_a({\bf r})$, 
Eq.(\ref{defphi}), are defined by choosing the radial part as
\begin{equation}
u_a(r) = r^{n_a-l_a-1} e^{-\alpha r}
\end{equation}
where $n_a=1,2,...$ is the principal quantum number and $l_a$ the total angular 
momentum
\begin{equation}
l_a=a_x+a_y+a_z.
\end{equation}
In what follows, we shall employ the usual notation for STO orbitals, namely 
$1s=e^{-\alpha r}, 2s=r e^{-\alpha r},
3s=r^2 e^{-\alpha r}, 2p_x= x e^{-\alpha r}, 3p_x=x r e^{-\alpha r}, 3d_{xx}= x^2 e^{-\alpha r}$,
 and so on. 

For the particular case of STO orbitals, we have not built the 
gaussian approximations of the radial part, Eq.(\ref{defua}) by minimization of the $\chi^2$, Eq.(\ref{chi2}). Instead, we have 
preferred to use the accurate representations of the exponential
\begin{equation}
e^{-r}=\sum_{i=1}^{n_g} c_i e^{-\gamma_i r^2}
\label{defexp}
\end{equation}
given by Lopez {\it et al.}\cite{integrals} for
a number of gaussian functions ranging from $n_g=1$ to $n_g=30$.
For $n=2$, $r^{n-1} e^{-r}$ is expanded as 
\begin{equation}
r e^{-r}=2\sum_{i=1}^{n_g} c_i \gamma_i r^2 e^{-\gamma_i r^2},
\label{expan}
\end{equation}
an expression obtained by considering the derivative
$\frac{\partial}{\partial a} e^{-ar}\Bigr\rvert_{a = 1}$ where the gaussian expansion of 
$e^{-ar}$, Eq.(\ref{defexp}), is used.
For $n=3$, the polynomial $r^2=x^2+y^2+z^2$ is withdrawn from the radial part and transferred 
to the polynomial part of the orbital.\\

Several choices of the functional form for the coordinate transformation have been investigated.
The following simple form is proposed
\begin{equation}
\tilde{\bf r}_i({\bf r}_1,{\bf r}_2) =
f(|{\bf r}_i|)
{\bf r}_i \;\;\; i=1,2.
\label{defv}
\end{equation}
where $f$ is a general (smooth enough) function.
In this case, the Jacobian is given by 
\begin{equation}
J({\bf r}_1,{\bf r}_2)=j(r_1)j(r_2)
\end{equation}
where
\begin{equation}
j(r) = f^2(r) |f(r)+r f'(r)|.
\end{equation}
Let us now use the ZV equation to get information on $f$.
Fixing electron 2 at some position ${\bf r}_2$, 
the ZV equation writes 
$$ 
f^2(r_1) |f(r_1)+r_1 f'(r_1)| \frac{1} {|{\bf u}_1(\tilde{\bf r}_1)-{\bf u}_2(\tilde{\bf r}_2)|}
$$
\begin{equation}
\times \frac{
\rho_{ab}[{\bf u}_1(\tilde{\bf r}_1)]\rho_{cd}[{\bf u}_2(\tilde{\bf r}_2)]-
\rho^G_{ab}[{\bf u}_1(\tilde{\bf r}_1))\rho_{cd}({\bf u}_2(\tilde{\bf r}_2)]}
{\pi_0({\bf r}_1,{\bf r}_2)}
= C
\end{equation}
where $C$ is a constant collecting the terms independent of ${\bf r}_1$. 
Without the coordinate transformation ($f=1$) the left-hand-side diverges 
in the large $r_1$-limit as the ratio
\begin{equation}
\sim \frac{
e^{-\frac{(\alpha+\beta)}{\sqrt{\zeta}} r_1}  }
{ e^{-\frac{ {r_1}^2 }{2} } }.
\end{equation}
Here, we have used the fact that
in the large-distance limit $\rho^G_{ab}\rho^G_{cd}$ 
becomes negligible with respect to $\rho_{ab}\rho_{cd}$. 
The divergence makes the variance of the estimator infinite and 
the Monte Carlo estimators do not converge (see, Fig.1 below).
Now, by using the coordinate transformation the divergence can be removed, for example 
by using the simplest form 
\begin{equation}
f(r) = \mu r^{\nu}
\label{deff}
\end{equation}
where 
\begin{equation}
\mu = \frac{\kappa \sqrt{\zeta} }{2 (\alpha+\beta)},
\label{defmu}
\end{equation}
$\kappa$ a positive constant and
$\nu$ some real exponent.
Now, taking expression (\ref{zeta}) for $\zeta$, the preceding ratio becomes
\begin{equation}
\sim \frac{ e^{   -\frac{\kappa}{2} r_1^{\nu +1}    }         }
          { e^{-\frac{{r_1}^2}{2}   }                  }
\end{equation}
The divergence is removed when
($\nu = 1$ and $\kappa \ge 1$) or ($\nu > 1$ and $\kappa \ge 0$).
In applications both parameters
can be optimized by minimization of the statistical fluctuations. Of course, more 
elaborate forms for $f$ can be used, this is let for future work.
\section{Numerical applications}
\label{num}
\subsection{Single representative two-electron integrals}
\subsubsection{Removing the infinite variance}
\begin{figure}[H]
\includegraphics[angle=0,width=1.1\columnwidth]{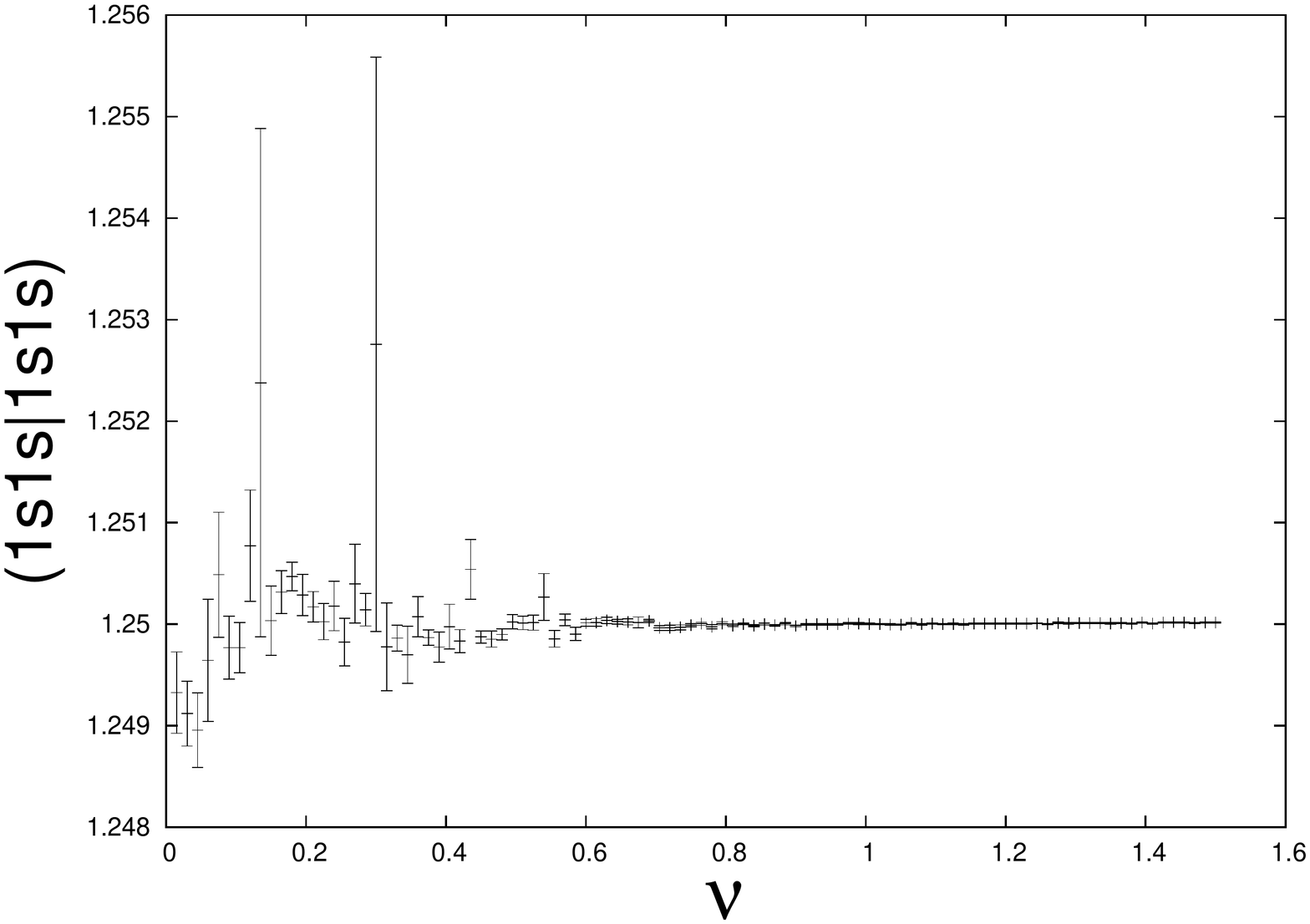}
\caption{Value of $I=(1s 1s|1s 1s)$ for $N=10^6$ and $n_g=5$ 
as a function of $\nu$. Coordinate transformation paramter, $\kappa=1$.}
\label{fig1}
\end{figure}

In the absence of the coordinate transformation we have seen that 
the Monte Carlo estimators of the STO integrals have an infinite variance. 
This point is illustrated in Figure \ref{fig1} where the Monte Carlo average 
as a function of the exponent $\nu$ 
of the function $f$ involved in the coordinate transformation [Eqs.(\ref{defv}) 
and (\ref{deff})] is shown. 
The results are presented for the simplest possible STO integral, namely
\begin{equation}
(1s1s|1s1s)= \frac{1}{(4\pi)^2}
\int d{\bf r}_1 d{\bf r}_2
e^{-r_1} \frac{1}{r_{12}} e^{-r_2}= \frac{5}{4}
\end{equation}
but a similar behavior is obtained for all STO integrals considered here.
The calculation is performed by approximating the $1s$ orbital with five gaussian functions, 
$n_g=5$, Eq.(\ref{defexp}), and by drawing $N= 10^6$ Monte Carlo configurations. The constant 
$\kappa$ in $f$, Eqs.[\ref{defmu}], is taken to be equal to 1.
As expected, for small values of $\nu$ 
uncontrolled fluctuations resulting from the infinite variance are present.
For large enough $\nu$ the wild fluctuations disappear and 
the Monte Carlo average becomes very close to the exact value of 1.25. 
The critical value of $\nu$ corresponding to the change of regime 
is compatible with $\nu \apprge 1$.

\subsubsection{ $(1s_A 1s_B|1s_C 1s_D)$ for STO atomic orbitals}
As a first application, we consider the calculation of four-center 
two-electron integrals over $1s$ Slater-type orbitals. For quantitative 
comparison we calculate the four integrals introduced by P\'erez {\it et al.}\cite{integrals} and 
presented in Tables I-IV of their work.
Following their convention for the normalization constant, the $1s$ orbital is written as
\begin{equation}
1s_A({\bf r}) =\mathcal{N}_{\alpha} e^{-\alpha|{\bf r}-{\bf A}|}.
\end{equation}
with 
\begin{equation}
\mathcal{N}_{\alpha}=\sqrt{\frac{\alpha^3}{\pi}}.
\end{equation}
The four two-electron integrals are denoted here as $I_k$ with $k$ ranging from 1 to 4; 
the exponents and nuclei positions are given in Tables I-IV of [\onlinecite{integrals}]. 

In Table \ref{tab1} the convergence of the two-electron integral $I_1$ as a function of the 
number of Monte Carlo drawings $N$ and number of gaussian functions $n_g$ is presented. 
The parameters of the coordinate transformation are taken to be $\kappa=1$ and $\nu=1$.
All computed values are in agreement with the exact value 
within the 2-$\sigma$ limits of the confidence interval. Here, the exact value is evaluated  
by using the approximate gaussian integral with the most 
accurate $n_g$=30-representation of the exponential function to our disposal, that is
$I^G(n_g=30)=0.1426742806$. All given digits are converged as a function of $n_g$ and 
the value is in very close agreement with that given in [\onlinecite{integrals}],
$I=0.14267429$ (difference of about $10^{-8}$).

As it should be, the statistical error decreases both as a function of $N$ 
at fixed $n_g$ and of $n_g$ at fixed $N$. At fixed $n_g$, the error decreases 
as $\sim \frac{1}{\sqrt{N}}$ as expected in a Monte Carlo calculation. 
When passing from $n_g$ to $n_g+1$ at fixed $N$, an average reduction of the statistical error
between 2 and 3 is obtained, except for $n_g=1$ where the factor is about 10, 
a larger reduction resulting from the very poor representation of the radial part 
$u(r)$ using only a single gaussian function. The gain in accuracy 
when increasing $n_g$ being directly related to the quality of 
the fit, no general rule is expected for it as a function of $n_g$. 
For each $n_g$ the value of the approximate gaussian integral
$I^G_1=(1s^G_A 1s^G_B|1s^G_C 1s^G_D)$ is also reported. These values allow 
to quantify the magnitude of the bias $\epsilon=|I_1-I^G_1|$ recovered by the Monte Carlo part. 
Of course, the approach is of interest only if the statistical error on the 
unbiased ZVMC integral is smaller than $\epsilon$.
Table \ref{tab1} shows that it is always the case, 
except for the smallest number of Monte Carlo steps $N=10^3$ (for almost all $n_g$) 
and also for $N=10^5$ with $n_g=7$.
The most accurate value of the integral is obtained for the largest value of $n_g$ and $N$ and 
is only in error of about
$9 \times 10^{-10}$. As we shall see below, such a typical accuracy will be sufficient to perform 
molecular calculations.

In Table \ref{tab2} the results for the three other two-electron integrals, $I_{k=2-4}$ 
as a function of $n_g$ and for $N=10^{11}$ are reported.
For $n_g$=7 the absolute errors on the integrals $I_2$, $I_3$, and $I_4$ are comparable 
to those obtained for $I_1$. The biases $\epsilon$ associated 
with the $n_g$=7-gaussian approximation 
are about $3 \times 10^{-7}$, $2 \times 10^{-5}$, and $3 \times 10^{-8}$ for $I_2$, $I_3$ and $I_4$, respectively. 
These biases are much larger than the corresponding statistical errors on the Monte Carlo values 
which are $4 \times 10^{-10}$, $2 \times 10^{-10}$, and $2 \times 10^{-11}$, respectively. 
It illustrates the effectiveness 
of the Monte Carlo approach for recovering the exact STO values, starting from
the approximate gaussian ones.

\begin{table*}
\caption{Convergence of the two-electron integral $I_1$ as a function of the number 
of Monte Carlo drawings $N$ and gaussian functions $n_g$. Error bars on the last digit (one-sigma confidence intervals) given in parentheses. $I^G_1$ value of the approximate reference gaussian integral. Parameters 
of the coordinate transformation ($\kappa=1$, $\nu=1$). Exact value obtained with $n_g=30$, see text.}
\begin{tabular}{lllllll}
\hline
$n_g \;$($I^G_1$)      &$N=10^3$   &$N=10^5$    &$N=10^7$& $N=10^9$     & $N=10^{11}$ \\
\\
1 $\;$(0.01982738)&0.138(5)   &0.1419(6)   &0.14272(6)   & 0.14267(1)   & 0.142674(1) \\
2 $\;$(0.14202888)&0.1420(6)  &0.14273(6)  &0.142659(5)  & 0.142674(1)  & 0.1426744(1)  \\
3 $\;$(0.14251733)&0.1424(2)  &0.14266(2)  &0.142679(2)  & 0.1426743(4) & 0.14267428(4)\\
4 $\;$(0.14269676)&0.14275(8) &0.142681(6) &0.1426741(6) & 0.1426743(2) & 0.14267428(1)\\
5 $\;$(0.14266530)&0.14265(3) &0.142674(2) &0.1426741(2) & 0.14267429(6)& 0.142674275(5)\\
6 $\;$(0.14267538)&0.142674(8)&0.142676(1) &0.1426743(1) & 0.14267428(2)& 0.142674282(2) \\
7 $\;$(0.14267435)&0.142680(4)&0.1426743(5)&0.14267419(5)& 0.14267428(1)& 0.1426742810(9) \\
\\
$I^{exact}_1= 0.1426742806$  & &            &             &        \\
\\
\hline
\end{tabular}
\label{tab1}
\end{table*}

\begin{table*}
\caption{Two-electron integrals $I_2, I_3$, and $I_4$ as a function of the number of 
gaussian functions $n_g$ with $N=10^{11}$. Error bars on the last digit given in parentheses. 
$I^G_k$, values of the approximate gaussian integral. Coordinate transformation parameters 
($\kappa=1$, $\nu=1$). $I^G_k(n_g=30)$ used as exact values for the integrals.
}
\begin{tabular}{llll}
\hline
$n_g$&$(\;\;\;\;\;\;\;\;I^G_2\;\;\;\;\;\;\;)\;\;\;\;\;\;\;\;\;I_2$ 
     &$(\;\;\;\;\;\;\;\;I^G_3\;\;\;\;\;\;\;)\;\;\;\;\;\;\;\;\;I_3$
     &$(\;\;\;\;\;\;\;\;I^G_4\;\;\;\;\;\;\;)\;\;\;\;\;\;\;\;\;I_4$\\ 
\\
4 &(0.030664621) 0.030682219(6) &(0.025796634) 0.026031858(6)&(0.6215 $10^{-6}$) 1.3298(3)  $10^{-6}$\\
5 &(0.030685617) 0.030682223(2) &(0.025932666) 0.026031864(3)&(0.9703 $10^{-6}$) 1.3299(2)  $10^{-6}$\\ 
6 &(0.030681829) 0.0306822243(9)&(0.025991113) 0.026031866(1)&(1.1750 $10^{-6}$) 1.32999(9) $10^{-6}$\\
7 &(0.030681940) 0.0306822237(4)&(0.026015525) 0.0260318663(7)&(1.2711 $10^{-6}$) 1.329997(4)$10^{-6}$\\ 
\\
  &$\;\;\;\;\;\;\;\;\;\;\;\;I^{exact}_2$=0.0306822234$^a$ & $\;\;\;\;\;\;\;\;\;\;\;\;I^{exact}_3$
=0.0260318660$^a$ & $\;\;\;\;\;\;\;\;\;\;\;I^{exact}_4$=1.330001 ${10^{-6}}^a$\\
\\
\hline
\end{tabular}
\caption*{
$^a$ Exact values reported in [\onlinecite{integrals}]: $I^{exact}_2$=0.03068223, $I^{exact}_3$=0.02603187, and $I^{exact}_4$=1.330001 $10^{-6}$.}
\label{tab2}
\end{table*}

\subsubsection{$(n_A l_A n_B l_B|n_C l_C n_D l_D)$ for STO atomic orbitals}
In Table \ref{tab3} some results for four-center two-electron integrals over STO orbitals 
with non-zero angular momenta are presented. The atomic orbitals considered are 
$1s_A=\mathcal{N}_{\alpha} e^{-\alpha |{\bf r}-{\bf A}|}$, 
$2p_A=\mathcal{N}_{\alpha} (x-x_A)e^{-\alpha |{\bf r}-{\bf A}|}$, 
and $3d_A= \mathcal{N}_{\alpha}(x-x_A)^2 e^{-\alpha |{\bf r}-{\bf A}|}$, 
with the same choice for the other nuclei. Ten particular integrals combining these
atomic orbitals have been selected. 
The nucleus centers and exponents have been chosen to avoid any particular spatial symmetry.
$I^G(n_g=30)$ are taken as exact values, all reported digits being converged as a function 
of $n_g$. For a given value of $n_g$, all ten integrals 
are calculated simultaneously over the same Monte Carlo configurations. Most of the 
computational effort is spent in computing quantities independent of the polynomial part 
of the orbitals, Eq.(\ref{defphi}). Thus,
the additional cost for calculating all integrals compared to that needed for the 
$(1s_A 1s_B |1s_C 1s_D)$ integral alone is marginal. It is one of the 
attractive properties of the approach. For $n_g=7$ the absolute 
errors obtained for the ten integrals range from $7 \times 10^{-9}$ to $5 \times 10^{-8}$. 
For a fixed number of drawings, the statistical error is proportional 
to the square root of the variance of the estimator. 
As the total angular momentum $L=l_A+l_B+l_C+l_D$
is increased, the variance is also expected to increase (higher and higher moments of 
the probability distribution are calculated) and so the error.
It is indeed what is observed in Table \ref{tab3} where the error 
increases continuously when going from $L=0$ to $L=5$. However, 
the absolute errors obtained in the less favorable case ($L=5$) are still very small.
\begin{table*}
\caption{
Two-electron integrals for STO orbitals with non-zero momenta. 
$\alpha=1,\beta=1.2,\gamma=1.6, \delta=2.1$;
${\bf A}$=(0.4,-0.2,0.5), ${\bf B}$=(-0.5, 0.3,-0.4), ${\bf C}$=(0.5,-0.6,0.6), ${\bf D}$=(-0.4,0.5,-0.4), $N=10^{11}$. Coordinate transformation parameters
($\kappa=1$, $\nu=1$). Exact values of the integrals obtained with $n_g=30$ (all digits converged).
}
\begin{tabular}{cclllll}
\hline
$I$& $L=\sum_M l_M$ &$\;\;\;\;\;\;\;\;n_g=4$&$\;\;\;\;\;\;\;\;n_g=5$&$\;\;\;\;\;\;\;\;n_g=6$&$\;\;\;\;\;\;\;\;n_g=7$&$\;\;\;\;\;\;\;\;$$I_{ex}$\\
\\
$(1s_A 1s_B|1s_C 1s_D)$& 0& \; 0.15920106(2)& \;0.159201058(6)& \; 0.159201059(3)& \; 0.159201062(1) & \;0.1592010625\\
$(2p_A 1s_B|1s_C 1s_D)$& 1&-0.07740410(1)&-0.077404130(6)& -0.077404122(2)& -0.077404126(1) &-0.0774041258\\
$(2p_A 2p_B|1s_C 1s_D)$&2& \;  0.07231812(3)& \;0.07231812(1) & \; 0.072318124(4)&  \;0.072318123(2) & \;0.0723181226\\
$(3d_A 1s_B|1s_C 1s_D)$&2&  \; 0.14198182(3)& \;0.14198184(1) & \; 0.141981835(5)&  \;0.141981837(2) & \;0.1419818359\\
$(2p_A 1s_B|2p_C 1s_D)$&2& \;  0.05575257(2)& \;0.055752572(7)& \; 0.055752568(3)&  \;0.055752573(1) & \;0.0557525723\\
$(2p_A 2p_B|2p_C 1s_D)$&3& -0.03943272(3)&-0.03943274(1) & -0.039432728(4)& -0.039432728(2) &-0.0394327283\\
$(3d_A 1s_B|2p_C 1s_D)$&3& -0.08961003(4)&-0.08961006(2) & -0.089610040(6)& -0.089610044(2) &-0.0896100435\\
$(2p_A 2p_B|2p_C 2p_D)$&4&  \; 0.01980999(5)& \;0.01980997(2) & \; 0.019809984(8)&  \;0.019809988(3) & \;0.0198099811\\
$(3d_A 1s_B|2p_C 2p_D)$&4&  \; 0.03393439(6)&\; 0.03393439(2) & \; 0.033934397(8)& \; 0.033934403(3) & \;0.033934395\\
$(3d_A 1s_B|3d_C 2p_D)$&5& -0.0386191(1) &-0.03861923(6) & -0.03861921(2) & -0.03861922(1)  & -0.038619232\\
\hline
\end{tabular}
\label{tab3}
\end{table*}

\subsection{Application to atomic and molecular systems}
In this section Hartree-Fock and near full CI calculations
using Slater-type atomic orbitals for Be, CH$_4$, and [H$_2$N(CH)NH$_2$]$^+$ are presented. 
For that, the full set of two-electron integrals
is to be computed. After removal of the redundancy among
orbital indices the number of integrals is about
$\frac{N_b^4}{8}$, where $N$ is the number
of orbitals (basis functions). The sampling distribution being independent of the orbitals,
all integrals are computed over the same Monte Carlo realization. In this way, the approach
is embarrassingly parallel not only under splitting of the full set of integrals
into independent blocks as in any approach but also with respect to the Monte Carlo sampling
that can be performed on independent blocks over an arbitrary number of compute cores.
The CI calculations are performed using the 
CIPSI algorithm\cite{Huron_1973}(Configuration Interaction using a 
Perturbative Selection made Iteratively) as implemented in the freely available 
electronic structure software QUANTUM PACKAGE.\cite{qp2} 
CIPSI combines a selected CI (sCI) step based on a second-order energetic criterion 
to select perturbatively the most important determinants 
and on a perturbative step where the second-order Epstein-Nesbet
pertubative estimate $E_{PT2}$ of the difference between the FCI 
and the variational reference energy is evaluated. $E_{PT2}$ is efficiently 
computed with a recently proposed hybrid stochastic-deterministic algorithm.\cite{pt2stoc}
In order to extrapolate the sCI results to the FCI limit, 
the method recently proposed by Holmes, Umrigar and Sharma
in the context of the HBCI method\cite{sharma} is employed.
It consists in extrapolating the sCI energy, $E_{sCI}$, 
as a function of $E_{PT2}$, 
{\it i.e} $E_{sCI} \simeq E_{FCI}-E_{PT2}$. 
When $E_{PT2}=0$, the FCI limit
has effectively been reached. This extrapolation procedure
has been shown to be robust, even for challenging chemical situations.
In the calculations presented here the number of selected determinants
is about a few millions and E$_{PT2}$ is small enough to enter the quasi-linear regime of
the difference $E_{FCI}-E_{sCI}$ as a function of the number of selected determinants.
Our estimate of FCI is denoted as exFCI (extrapolated FCI).
For the various aspects of the CIPSI implementation and several examples molecular applications 
the interested reader is referred to [\onlinecite{qp2}] and references therein.

Very few STO basis sets adapted to post-HF
calculations (that is, including optimized polarization functions to describe the virtual space) 
have been proposed in the literature.
Here, in all applications we employ the Slater-type atomic orbital (STO) 
valence basis set VB1 developed by Ema {\it et al.}\cite{ema} for the first and second row
atoms. 

\subsubsection{Beryllium atom}
For Be the VB1 basis set consists of two $1s$, three $2s$ and one $2p$, 
for a total of 8 atomic STO basis functions and about 700 
two-electron integrals to evaluate. The second column of table \ref{be_scf} gives 
for increasing values of $n_g$
the Hartree-Fock energies obtained with the gaussian basis sets used in the deterministic 
part of the calculation. Denoted here as $\{n_g\}$, 
these GTO basis sets are made of the approximate gaussian orbitals $\phi^G_a$, as expressed in 
Eq.(\ref{defphig}). By definition, they have the same size 
as the STO basis set, they only differ by the quality of the approximation made for 
representing the STO orbitals.
As it should be, as $n_g$ increases the Hartree-Fock energies 
converge to the exact Slater Hartree-Fock energy of -$14.572976251$. This latter value has been 
computed using the exact expressions for the one- and two-electron STO integrals that are known 
in the case of a single nucleus center.
Note that this value is in perfect agreement with that given by Ema {\it et al.}\cite{ema}
The third column gives the Hartree-Fock energies obtained with the STO integrals 
computed with ZVMC and using the $\{n_g\}$ gaussian basis set for the deterministic part.
The number of Monte Carlo drawings is $N=10^7$ and 
the coordinate transformation parameters ($\kappa=3.2$, $\nu=1$). The value of $\kappa$ 
has been optimized by minimization of the statistical 
error. As it should be, the HF energies obtained with the STO integrals computed by ZVMC 
are independent of the $n_g$-approximation, only the magnitude of the statistical error is affected.
This error decreases very rapidly as a function of $n_g$ ranging from $10^{-5}$ a.u. to less than 
$10^{-8}$. For $n_g=14$ the value of -14.57297625(1) is in
perfect agreement with the exact STO Hartree-Fock energy.

In Table \ref{be_fci} the exFCI values for Be are presented. 
For comparison the exFCI value computed with 
the exact one- and two-electron STO integrals and the FCI value of Ema {\it et al.} are given. 
Similarly to the Hartree-Fock results, i.) the exFCI values obtained with the $\{n_g\}$ gaussian 
basis sets converge to the exact ones as $n_g$ increases, ii.)
the exFCI values obtained with the ZVMC STO integrals are independent of $n_g$,  
and iii.) the statistical error decreases rapidly as the gaussian approximation is improved.
For $n_g=10$ our exFCI energy is converged 
with 7 decimal places and is in full agreement with the exact value.

\begin{table*}
\caption{Be atom. Hartree-Fock (HF) energies for the approximate gaussian basis sets 
$\{n_g\}$ (see, text) and for the exact STO basis set with Monte Carlo integrals computed 
using the $\{n_g\}$ basis set for the deterministic part.
Total statistics: $N=10^7$, coordinate transformation parameters ($\kappa=3.2$, $\nu=1$). 
Energies in atomic units.}
\begin{tabular}{lll}
\hline
$n_g$ & $E_{HF}(\{n_g\}$)  & $E_{HF}$(STO) using $\{n_g\}$ basis set\\
\\
4  & -14.53734687 &\;\;\;\;\;\; -14.57296476(1190)\\
6  & -14.56851725 &\;\;\;\;\;\; -14.57297487(166)\\
8  & -14.57230008 &\;\;\;\;\;\; -14.57297620(30)\\
10 & -14.57285628 &\;\;\;\;\;\; -14.57297620(9)\\
12 & -14.57295200 &\;\;\;\;\;\; -14.57297627(2)\\
14 & -14.57297086 &\;\;\;\;\;\; -14.57297625(1)\\
20 & -14.57297624 & $\;\;\;\;\;\;\;\;\;\;\;\;\;\;\;\;\;\;$-\\
\\
Exact Slater HF energy & -14.572976251 & \\
HF energy, Ema {\it et al.}\cite{ema}  & -14.572976 &\\
\hline
\end{tabular}
\label{be_scf}
\end{table*}

\begin{table*}
\caption{Be atom. exFCI energies for the approximate gaussian basis sets
$\{n_g\}$ (see, text) and for the exact STO basis set with Monte Carlo integrals computed
using the $\{n_g\}$ basis set for the deterministic part.
Total statistics: $N=10^7$, coordinate transformation parameters ($\kappa=3.2$, $\nu=1$).
Energies in 
atomic units.}
\begin{tabular}{lll}
\hline
$n_g$ & $E_{exFCI}(\{n_g\}$) & $E_{exFCI}$(STO) using $\{n_g\}$ basis set\\
\\
4 &        -14.58193749 &\;\;\;\;\;\; -14.6179914(114)\\ 
6 &        -14.61342650 &\;\;\;\;\;\; -14.6180006(19)\\ 
8 &        -14.61730631 &\;\;\;\;\;\; -14.6180019(3)\\
10 &       -14.61787858 &\;\;\;\;\;\; -14.6180020(1)\\
\\
Slater exFCI energy & -14.61800193 & \\
FCI energy, Ema {\it et al.}\cite{ema}  & -14.618002 & \\
\hline
\end{tabular}
\label{be_fci}
\end{table*}

\subsubsection{CH$_4$}
\begin{figure}
\begin{center}
\includegraphics[angle=0,width=1.1\columnwidth]{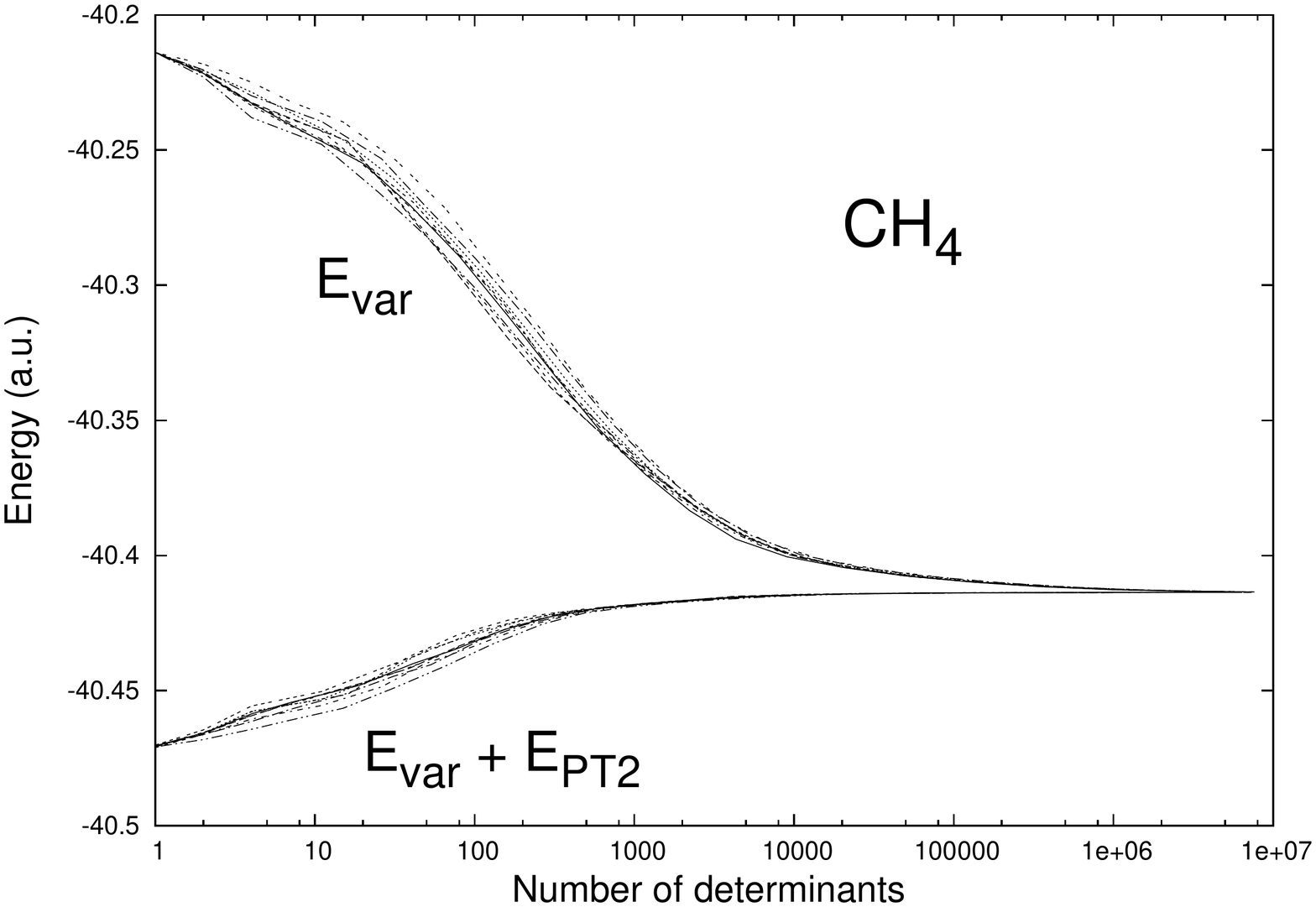}
\caption{CH$_4$ molecule. Convergence of the CIPSI variational energy $E_{var}$ (upper curve) and 
$E_{var}$ + $E_{PT2}$ energy (lower curve) as a function of the number of selected determinants.
Convergence curves realized for 10 statistically independent ZVMC calculations of the 
two-electron STO integrals. Energies in atomic units.
}
\label{fig_ch4}
\end{center}
\end{figure}

Table \ref{ch4} presents the HF and $1s^2$-frozen-core exFCI energies of CH$_4$. The 
geometry of the molecule -close to the experimental one- is given in the Supporting Information. 
Results using the $\{$6-9$\}$ and VB1 STO basis sets are shown. For comparison,
we also report those obtained with the cc-pVDZ basis set. Here, 
the notation $\{$6-9$\}$ refers to the gaussian 
basis set defined above, except that a different number of gaussian functions 
is used to approximate the various STO orbital,
the motivation being to get a more
uniform quality among orbital approximations.
To be more precise, the orbitals 
are generically expanded with $n_g=6$. For 2p and 3d orbitals, 
$n_g$ is increased by one unit ($n_g=7$) or two ($n_g=8$), respectively. In addition, 
when the exponent is too large (here, greater than 4)
$n_g$ is increased by 3 ($n_g=9$). 
The VB1 STO basis set is made of three 1s and one 2p for H and two 1s, four 2s, 
three 3p and one 3d, for a total of 44 
cartesian STO orbitals. The cc-pVDZ basis is made of 
35 cartesian orbitals. The number of two-electron integrals calculated with ZVMC 
is about 500 000. All mono-center integrals have been computed with the exact expressions for 
these STO integrals.
The total number of Monte Carlo drawings is about $3 \times 10^{11}$
The HF and FCI results are presented in table \ref{ch4}. Quite remarkably,
both the STO VB1 Hartree-Fock and exFCI energies are obtained with a very good accuracy, 
that is 6 and 5 decimal places, respectively. 
The statistical errors have been obtained by running ten statistically independent 
calculations. Figure \ref{fig_ch4} presents the convergence of the CIPSI energies as a function 
of the number of selected determinants up to $N_{det}=8 \times 10^{6}$ by plotting 
the ten curves obtained for each independent Monte Carlo run.
The upper curve shows the convergence of the CIPSI variational energy and the lower one 
the variational + $E_{PT2}$ energy. As seen the dispersion of the curves decreases as a function 
of the determinant and converge to the FCI energy. 

\begin{table*}
\caption{CH$_4$ molecule. Hartree-Fock and $1s^2$-frozen-core exFCI energies for the cc-pVDZ, 
$\{$6-9$\}$ (see, text), and Slater VB1 basis sets. 
Total statistics: $N \sim 3 \times 10^{11}$. Coordinate transformation parameters ($\kappa=3.2$, $\nu=1$).
Energies in atomic units.}
\begin{tabular}{lll}
\hline
Basis set & $E_{HF}$  & $E_{exFCI}$\\
\\
cc-pVDZ (GTO basis)  & -40.198743  & -40.392975\\ 
$\{$6-9$\}$ (GTO basis)    & -40.212117  & -40.410443\\
VB1 (STO basis)   & -40.21485042(7)     & -40.413651(3)\\
\hline
\end{tabular}
\label{ch4}
\end{table*}

\subsubsection{A simple cyanine: [H$_2$N(CH)NH$_2$]$^+$}

In the last application the results obtained for a simple model of cyanine molecule,
[H$_2$N(CH)NH$_2$]$^+$ are presented. 
The geometry of the molecule is available in the Supporting Information.
The VB1 basis set consists of 
(three 1s, one 2p) for H and (two 1s, three 2s, three 3p and one 3d) for C and N, 
for a total of 90 cartesian STO orbitals.
The total number of two-electron integrals is about 
8.3 $10^6$. Table \ref{cyanine} presents the Hartree-Fock and $1s^2$-frozen core exFCI results.
As seen the accuracy reached is lower than in the case of CH$_4$ but still 
very good. The statistical error on the exFCI energy is $2 \times 10^{-4}$ a.u. ($\sim$ 0.1 kcal/mole), that is, 
the sub-chemical accuracy is reached.

\begin{table*}
\caption{Cyanine molecule. Hartree-Fock and $1s^2$-frozen core exFCI results
for the cc-pVDZ,
$\{$6-9$\}$ (see, text), and Slater VB1 basis sets.
Total statistics: $N \sim 10^{10}$. Coordinate transformation parameters ($\kappa=3.6$, $\nu=1$).
Energies in atomic units.}
\begin{tabular}{lll}
\hline
Basis set & $E_{HF}$& $E_{exFCI}$\\
\\
cc-pVDZ (GTO basis) & -149.485140 & -150.0052\\
$\{$6-9$\}$ (GTO basis)  & -149.521267 & -150.0635(1)\\
VB1 (STO basis)  & -149.5298389(9) & -150.0738(2) \\
\hline
\end{tabular}
\label{cyanine}
\end{table*}

\section{Summary and discussion}
\label{conclusion}
In this work an efficient Monte Carlo approach to calculate general two-electron integrals 
has been presented. Using variance reduction techniques it has been shown that
the very high level of precision required on two-electron integrals by 
Hartree-Fock and post-HF calculations can be achieved. 

The major advantage of the approach is its great generality and flexibility.
It can be used with any type of orbitals provided that 
sufficiently accurate gaussian approximations are available for them. 
Various schemes can be used to construct such approximations, so 
it is not a severe practical limitation. Actually, the key point 
is that ZVMC results do not depend on this approximation (whatever its quality), 
only the magnitude of the statistical error is affected. We also note that the approach 
can be generalized without difficulty to various situations, 
for example, in the case of an arbitrary two-body interaction 
or for the calculation of three-particle integrals, the sole condition being that 
the approximate gaussian integrals involved can be efficiently evaluated.

Now, it is clear that the major drawback of the approach is its (very) 
high computational cost. In the applications presented in this work, 
the number of Monte Carlo drawings required to make molecular calculations possible 
ranges from 10$^9$ to 10$^{11}$. In terms of computational burden,
the most extensive simulation realized here (90 orbitals and about 8 millions
STO-type two-electron integrals for the cyanine molecule) has been performed 
using 4800 compute cores running in parallel during a few hours. 
Although such a cost may appear very high, we emphasize that
using the approach in the present form Monte Carlo calculations are feasible with 
the required accuracy and that we have already been able 
to realize for a system of the size of the cyanine molecule (24 electrons)
a FCI calculation involving 18 active electrons distributed among 90 orbitals,
which is, to the best of our knowledge, the most extensive molecular calculation 
performed so far using pure STO orbitals (no gaussian approximation, even for the challenging 
four-center two-electron integrals). However, there is clearly much room for the improvement of 
the method and 
(much) smaller timings should be easily reachable. Indeed, no particular attention has been 
paid here to the algorithmic implementation, our objective being mainly to demonstrate 
the feasibility of the method. No doubt that, in view of the simplicity of the approach
and the very repetitive character of the basic floating point operations to be performed, 
more efficient implementations taking full advantage of the most advanced capabilities 
of modern processors should be possible.
Furthermore, a lot remains to be done to improve the approach itself, particularly
in the way the correlated part is performed and in the choice of the coordinate transformation.
Finally, we would like to insist on the fact that the most interesting source of (indirect)
computational savings is the possibility of using more compact and physically meaningful 
basis sets, a key aspect considering the sharp increase of the 
cost of post-Hartree-Fock methods with the number of orbitals.

\begin{acknowledgements}
The author would like to thank P.F. Loos and A. Scemama for helpful discussions and a 
careful reading of the manuscript. We also thank
the Centre National de la Recherche Scientifique (CNRS) for funding. This work was
performed using HPC resources from i) GENCI-TGCC (Grant
No. 2018-A0040801738), and ii) CALMIP (Toulouse) under
allocations 2018-0510, 2018-18005 and 2019-18005.
\end{acknowledgements}

\bibliography{p}

\end{document}